\documentclass[pre,floatfix,twocolumn,aps,showpacs,superscriptaddress]{revtex4-1}
\usepackage{graphicx}
\usepackage{amsmath}
\usepackage{amssymb}
\usepackage{color}

\newcommand{\beq}{\begin{equation}}
\newcommand{\eeq}{\end{equation}}

\begin{document}
	
\title{Profusion of Transition Pathways for Interacting Hysterons}

\author{Martin van Hecke}
\affiliation{AMOLF, Science Park 104, 1098 XG Amsterdam, Netherlands}
\affiliation{Huygens-Kamerlingh Onnes Lab, Universiteit Leiden, P.O.~Box~9504, NL-2300 RA Leiden, Netherlands}
	
\date{\today}
	
\begin{abstract}
The response, pathways and memory effects
of cyclically driven complex media can be captured by
hysteretic elements called hysterons. Here we demonstrate the profound impact of hysteron interactions on pathways and memory.
Specifically, while the Preisach model of independent hysterons features a restricted class of pathways which always satisfy return point memory,
we show that three interacting hysterons generate more than 15,000 transition graphs, with most violating return point memory and having features
completely distinct from the Preisach model.
Exploring these opens a route to designer pathways and information processing in complex matter.
\end{abstract}
	
	
\maketitle


A rugged landscape governs the intermittent response of complex materials, which features sequences of transitions between metastable states that constitute pathways \cite{
Kramer,Cerda,Katan,Witten,Aharoni,Oppenheimer,Lahini,Bense,Murphy,Candelier,Losert,Bonn,Preisach,Barker,Middleton,Sethna,Crackle,Keim2}. Under cyclical driving, these pathways form hysteresis loops which may encode Return Point Memory (RPM), the widespread ability of complex materials to
revisit a previous state when the driving reaches a previous extremal value \cite{Preisach,RMP,MunganMert,Terzi,Barker,Middleton,Sethna}.
Microscopically, localized bistable entities such as spins or rearranging zones play a crucial role \cite{Preisach,RMP,MunganMert,Terzi,Regev,PreprintRegev,Bense}. These can be modeled as
{\em hysterons}, two-state elements which switch between phases ''0'' and ''1'' when the global driving field $U$ passes through the upper and lower ''bare'' switching fields $u_i^+$ or $u_i^-$ (Fig.~1) \cite{Preisach,RMP,Paulsen,MunganMert,Terzi,PreprintKaren,PreprintKeim,PreprintLindeman,Bense,Barker,Middleton,Sethna}.
Indeed,
collections of non-interacting hysterons---the well-studied Preisach model---describe surprisingly complex sequences of transitions and satisfy
 RMP \cite{MunganMert,Terzi,PreprintKaren,Bense,Barker,Middleton}.

Here we probe the rich physics of interacting hysterons. Interactions are physically expected \cite{Regev,Middleton,Bense,PreprintKeim,PreprintLindeman} and
while non-interacting hysterons switch their phases indepedently of each other, in sequences determined by the order of their bare switching fields, interactions scramble and entangle these switching orders.
To characterize the response of coupled hysterons
we use transition graphs (t-graphs), a recently introduced representation
that captures all pathways of a complex system and aids in
probing memory effects \cite{Paulsen,Regev,MunganMert,Terzi,PreprintKaren,PreprintKeim,PreprintLindeman,Bense,PreprintRegev}.

We investigate t-graphs of $n$ interacting hysterons, focussing on $n\!=\!2$ and $n\!=\!3$.
We show that interactions mushroom
the number of t-graphs, which feature transitions (avalanches, pseudo-avalanches, multi-edges) and topological structures (subharmonic cycles, breakdown of RPM) completely distinct from the Preisach model. Besides providing a fresh perspective on the response of complex media, we show that our study  paves the way for materials with designer pathways \cite{PreprintKaren,Bertoldi,Coulais,Mungan,PreprintZanaty}.

\begin{figure}[!t]
\begin{center}
\includegraphics[width=1\linewidth,bb = 90 192 730 350,clip
]{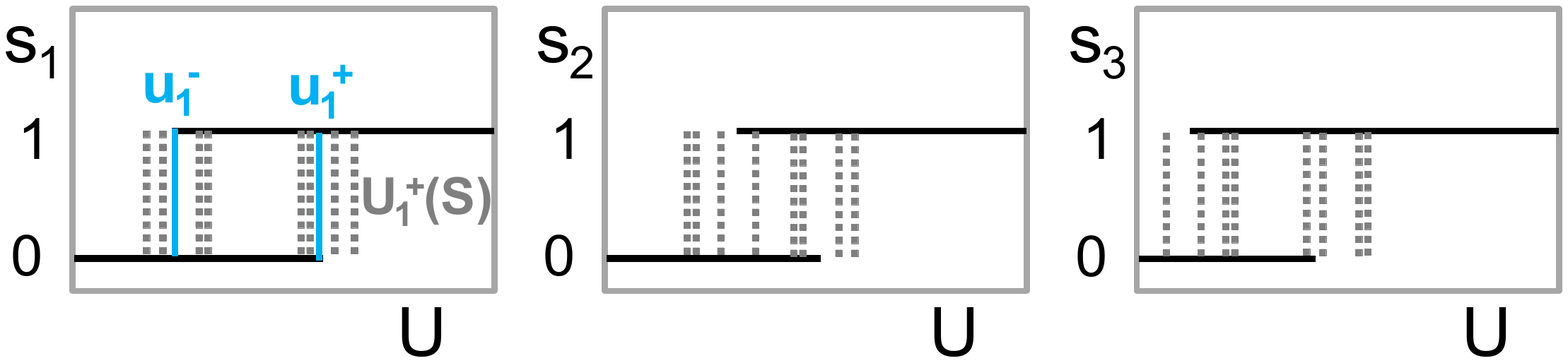}
\end{center}
\caption{(Color online)  Hysteresis diagrams for three interacting hysterons driven by a global field $U$. A single hysteron features internal phase $s_i=0,1$ and bare switching fields $u^{+,-}_i$ (blue, full line). Interactions lead to
multiple state-dependent switching fields
$U^{+,-}_i(S)$ (grey, dashed)  for each hysteron.
}
\label{fig:1}
\end{figure}


\section{Model and t-graph construction}

In this section we describe the model of linearly interacting hysterons in detail.
We first describe the models elementary state transitions (section \ref{sec:model}). We then introduce a recursive algorithm to obtain t-graphs for arbitrary realizations of the model (Section \label{sec:rec}), and for reference briefly show the t-graphs for the Preisach model (section \ref{sec:rec}).

\subsection{Model} \label{sec:model}

 We model interactions via
a linear dependence of the switching fields $U_i^\pm$ of
hysteron $i$ on the phases $s_j$ of all other hysterons  (Fig.~1):
\begin{equation}
U_i^{+,-}({S})=u_i^{+,-}-\Sigma_{j\ne i} ~c_{ij}s_j~,
\end{equation}
where $S$ denotes the state $ \{s_1,s_2,\dots\}$ and $u_i^{+,-}$ and
$c_{ij}$ are the bare switching fields and interactions constants
 \cite{PreprintKeim,PreprintLindeman}. We take $0\!<\!u_i^- \!<\!u_i^+\!<\!1$, require $u_1^+\!>\!u_2^+\!>\!\dots$ and   assume no degeneracies to occur
  \cite{MunganMert,Terzi}.
Furthermore, we choose a gauge where $c_{ii}=0$; note that
 nonzero diagonal interaction constants $c_{ii}$
only affect the values of $U_i^-$ and can readily be absorbed in the bare switching fields $u_i^-$.

The upper and lower switching fields for a given state $S$ follow from the minimum (maximum) value of the switching fields of each hysteron:
\begin{eqnarray}
U^+(S) &=& \min_{i_0} U_{i_0}^+(S)~,\\
U^-(S) &=& \max_{i_i} U_{i_1}^-(S)~,
\end{eqnarray}
where $i_0$ ($i_1$) runs over the hysterons that are 0 (1).
State $S$ becomes unstable when $U$ exceeds $U^+(S)$ or decreases below
$U^-(S)$, which initiates an up or down transition at critical driving value $U^c=U^\pm(S)$ (note that extremal states have only one transition).

\begin{figure}[!t]
\begin{center}
\includegraphics[width=1\linewidth,bb = 90 40 730 185,clip
]{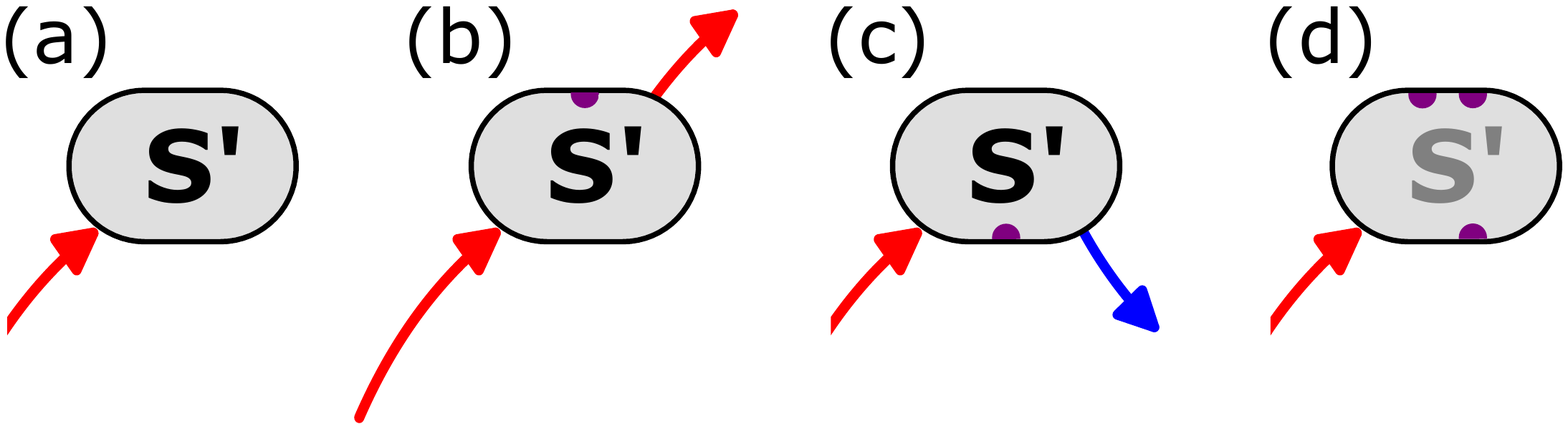}
\end{center}
\caption{(Color online)
Four scenario's after an up-transition reaches the state $S'$ at $U\!=\!U^c$ (red incoming arrow coming from below), depending on the stability of each hysteron in state $S'$ at $U\!=\!U^c$;
unstable hysterons in the 0 (1) phase are visualized as (purple) dots at the top (bottom) of the node $S'$.
(a) The transition terminates at $S'$ when all hysterons are stable at $U\!=\!U^c$.
(b) If one phase 0 hysteron  is unstable at $U\!=\!U^c$, this hysteron switches to phase 1, defining an additional step of the transition that now forms an avalanche
(outgoing arrow, red). (c) If one phase 1 hysteron is unstable at $U\!=\!U^c$,
this hysteron switches to phase 1, defining an additional step of the transition that now forms an avalanche (outgoing arrow, blue). (d) If more than one hysteron is unstable at $U=U^c$, we consider the model ill-defined. }
\label{fig:sce}
\end{figure}

Interactions can induce avalanches that proceed via intermediate, unstable states. For example, consider the case where $S$ initially transitions to a state $S'$
at value of $U$ where $S'$ is not stable, and then transitions to a stable landing state $S''$. We call this an avalanche of length two, denote it as $S\!\rightarrow\!S''$,
and ignore the intermediate states which in experiments would not be observable. By considering the stability of all hysterons in the transition state $S'$ at $U=U^c$, the following scenario's can arise (Fig.~\ref{fig:sce}).

(i) The transition state $S'$ is a stable landing state when $U^-(S')\!<\!U^c\!<\! U^+(S')$ yielding a transition between state $S$ and $S'$ (Fig.~\ref{fig:sce}(a)). While $S'$ is always stable at $U=U^c$ when $c_{ij}=0$,
interactions may cause state $S'$ to be unstable, leading to avalanches.

(ii) When one '0' hysteron of $S'$ is unstable, there is an additional up transition from $S'$ to $S''$ (Fig.~\ref{fig:sce}(b)), while when one '1' hysteron is unstable, there is an additional down transition from $S'$ to $S''$ (Fig.~\ref{fig:sce}(c)).
If state $S''$ is stable, we obtain an avalanche of length two between state
$S$ and $S''$ in which two hysterons change their phase at fixed $U\!=\!U^c$; if state $S''$ is unstable, we iterate and may find longer avalanches.

(iii) Occasionally more than one hysteron may be unstable in a given state (Fig.~\ref{fig:sce}(d)).
Despite not having been discussed before to the best of our knowledge, such scenario can readily be constructed. In fact, we have observed that such conditions can arise for arbitrarily weak interactions, and have a significant statistical weight for stronger interactions (see sections \ref{sec:II} and \ref{sec:III}). Hence, they are not a mathematical subtlety, but are an intrinsic feature of arbitrarily linearly coupled hysterons. We note that extensions of the model could be considered so that the 'most unstable' hysteron would switch and define state $S''$. However, here we choose to consider the corresponding t-graph to be  ill-defined whenever a state with multiple unstable hysterons arises.

\subsection{Recursive Algorithm and
Transition graphs}\label{sec:rec}
We collect all transitions between states that can be reached from the extremal states ${\bf 0}=\{0,0,\dots \}$ or ${\bf 1}=\{1,1,\dots \}$
into a t-graph via an iterative algorithm.
Starting from state $\{\bf{0}\}$, we determine its up transition and landing state,
and then iteratively determine the transitions from all fresh landing states, until no new fresh states can be found. The resulting t-graph contains all $N$ nodes that are reachable from the extremal states and $2N-2$ directed edges which represent the transitions, where each edge is labeled by its character (up or down), its critical field $U^+$ or $U^-$, and the number of intermediate steps.

We collect transitions between states that can be reached from the ground state ${\bf 0}=\{0,0,\dots \}$ into a t-graph via an iterative algorithm. This algorithm initiates a t-graph by node ${\bf 0}$ and  determines its up transition to a state $S_1$ following the procedure outlined above. Node $S_1$ is then added to the graph, as well as the edge ${\bf 0}\rightarrow S_1$.
Node ${\bf 0}$ is than labeled as 'stale', as all its transitions have been determined, and $S_1$ is labeled as a 'fresh' node. The up (provided $S_1\ne{\bf 1}=\{1,1,\dots\}$) and down transitions from the fresh state $S_1$ to states $S_2$ and $S_3$ are determined, state $S_1$ is labeled as stale, and states $S_2$ and $S_3$ are added to the t-graph and labeled as fresh provided they have not been visited before. This procedure is repeated until no more new fresh states are found.

Occasionally, loop-like avalanches of the form $S \rightarrow S' \dots \rightarrow S$ occur: we consider the corresponding t-graphs ill-defined. Again we note that such cases can readily constructed (we give an explicit example in section \ref{sec:IIdes}), and carry a significant weight. We suggest that more elaborate models, which for example would have an energy functional and dissipation, could be constructed to avoid such self-loops; for simple linearly coupled hysterons, they are an intrinsic feature of the model.

The resulting t-graph contains all $N$ nodes that are connected to state ${\bf 0}$ and the $2N-2$ directed edges which represents the transitions, where each edge is labeled by its character (up or down transition), its corresponding critical field $U^c$, and its length, i.e., the number of steps in an avalanche (one for an ordinary transition). Graphically, we order the nodes from bottom to top as function of their magnetization
$m=\Sigma s_i$, and from left to right lexicographically. Up and down transitions are colored red and blue, and the thickness of the edges represents the length of the avalanche.

T-graph construction involves evaluating ''design'' inequalities on the parameters
which govern each transition. These inequalities are not independent and vary strongly with the t-graph topology.
Moreover, certain parameters, in particular for strong interactions, may
yield avalanches that return to their initial state (self-loops), or contain states
where more than one hysteron is unstable; we consider the corresponding t-graphs ill-defined. Together, this makes
finding and classifying all t-graphs  complex.

\subsection{Physical Relevance}

The model that we discuss here is perhaps the simplest model in which hysterons are interacting, and as such is a natural choice, also studied in \cite{PreprintKeim,PreprintLindeman}. One might wonder if in physical systems, the interaction coefficients are free or are subject to additional constraints, such as reciprocity. While more work is needed, experiments on crumpled sheets  in which such interactions can be measured do not find any evidence for such constraints \cite{Bense,yoavpre}. In fact, these experiments suggest that the coupling coefficient $c_{ij}$ may take on different values---including with  different signs---for the upper and lower switching field, so that twice as much interactions coefficients might be needed to describe the full experiments \cite{Bense}. Nevertheless, the current model, despite having less free parameters, contains t-graphs of similar complexity as found experimentally. Finally, we stress that bistable elements are a work horse of mechanical metamaterials, and we are already finding that coupling these appropriately yields experimental realizations of interacting hysterons with accompanying complex t-graphs \cite{jerry}. Hence, while much more work is needed, the model studied here provides a solid jumping point for both experimental and  advanced theoretical studies.

	\begin{figure}[!t]
		\includegraphics[width=1\linewidth,bb = 000 000 198 132,clip]{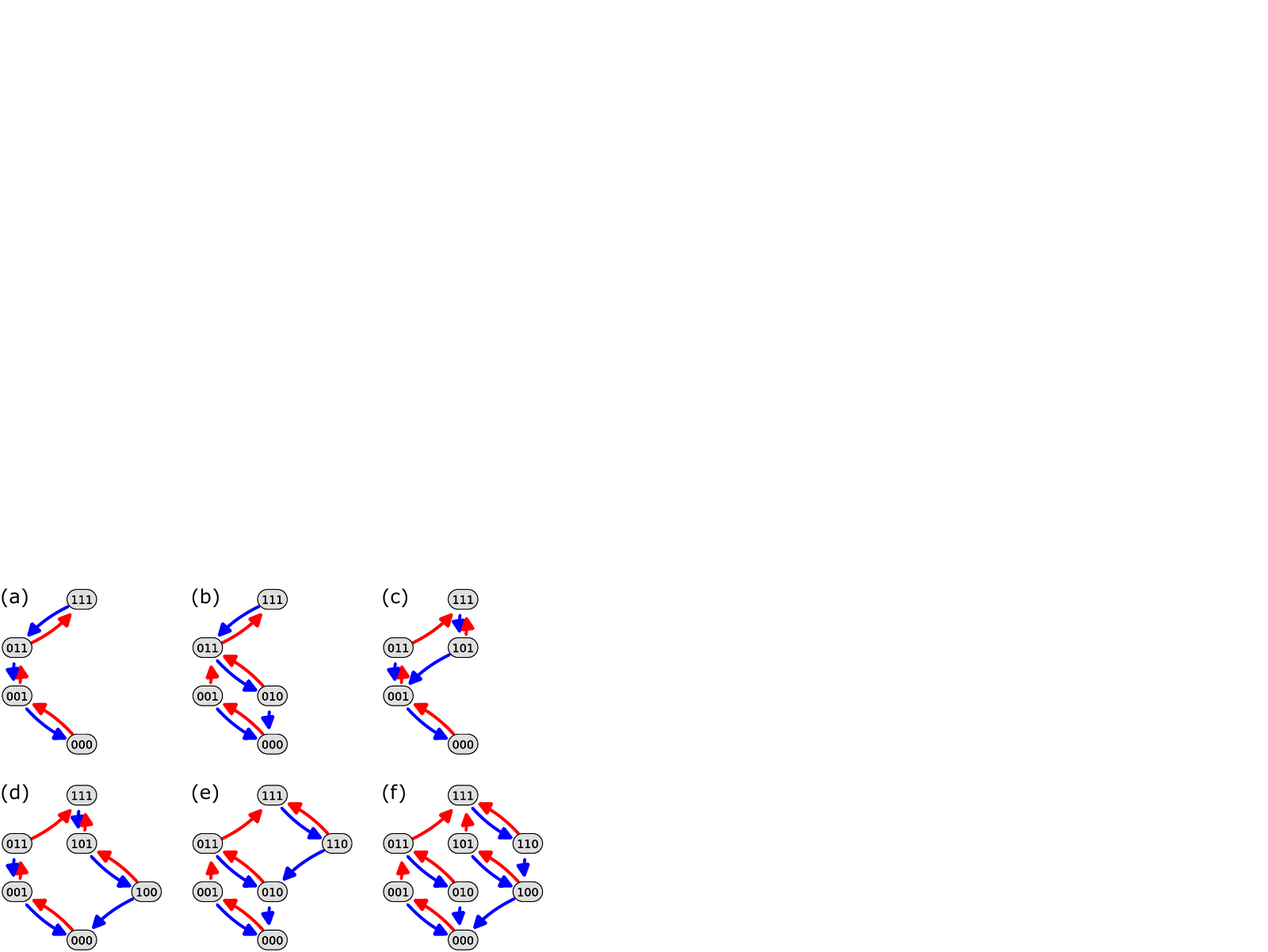}
		\caption{(Color online) T-graphs for the $n=3$ Preisach model (no interactions), for $u_1^+>u_2^+>u_3^+$.
Up and down transitions are represented by red (light grey)  and blue (dark grey) arrows respectively.
The ordering of the lower switching fields determines each graph \cite{MunganMert,Terzi}:
(a) $u_1^->u_2^->u_3^-$.
(b) $u_1^->u_3^->u_2^-$.
(c) $u_2^->u_1^->u_3^-$.
(d) $u_2^->u_3^->u_1^-$.
(e) $u_3^->u_1^->u_2^-$.
(f) $u_3^->u_2^->u_1^-$.
	 }\label{fig:p}
	\end{figure}

\subsubsection{Preisach t-graphs}

In the absence of interactions, the model is exactly the Preisach model, whose
t-graphs and properties have recently received renewed attention \cite{MunganMert,Terzi}.
Under the assumed ordering of the upper switching fields, each permutation of the
ordering of the lower switching fields  yields a unique t-graph, that we include
here (for $n\!=\!3$) to
gain familiarity with our representation of the t-graphs and for comparison
(Fig.~\ref{fig:p}). We note that the two $n=2$ Preisach t-graphs
are equivalent to the subgraphs of Fig.~2 obtained by pruning the $\{1xx\}$ states.

\begin{figure}[!t]
\begin{center}
\includegraphics[width= \columnwidth,bb = 0 00 312 161,clip]{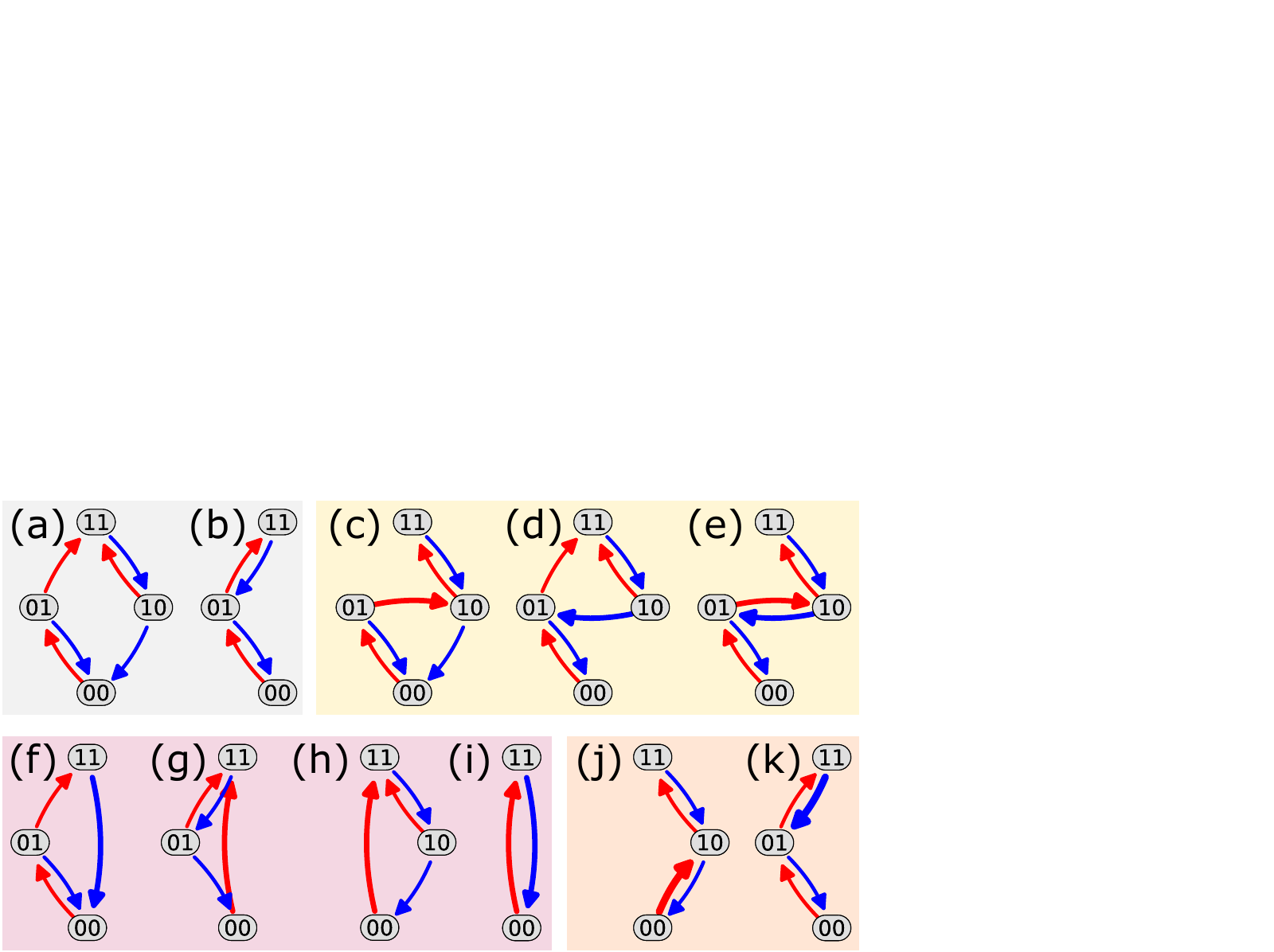}
\end{center}
\caption{(Color online) Distinct transition graphs for $n=2$ coupled hysterons. (a-b) For $c_{ij}=0$
we recover the well-known Preisach t-graphs \cite{MunganMert,Terzi}. (c-e) For $c_{ij} <   0$, 'horizontal'
avalanches of length two (thick arrows) may occur.
(f-i) For $c_{ij} > 0$, four additional t-graphs with vertical avalanches may occur. (j-k) For $c_{ij}$ of mixed sign, two additional t-graphs featuring pseudo-avalanches of length three (thick arrows) are observed.
}
\label{Fig:4}
\end{figure}

\section{Two interacting hysterons}\label{sec:II}
In this section, we determine all possible t-graphs for two interacting hysterons by exhaustively sampling the parameter space span by the
four bare switching fields $u_i^{+,-}$ and two coupling coefficients
$c_{12}$ and $c_{21}$ (section \ref{sec:IInovel}). We then work out the precise constraints on these parameters for each t-graph to occur (section \ref{sec:IIdes}).
Finally, we present their  statistical likelihood as function of interaction strength and uncover powerlaw scaling (section \ref{sec:IIpwl}).

Our results show that the underlying inequalities on the design parameters already become numerous and have an intricate structure for $n=2$, reflecting the complexity of the underlying problem. Crucially, our results clarify and quantify how even weak interactions lead to a significant growth in the number and variety of t-graphs.

\subsection{Novel t-graphs due to interactions}\label{sec:IInovel}

Without interactions, there are only two distinct t-graphs (Fig.~\ref{Fig:4}(a)-\ref{Fig:4}(b)), selected by the sign of $u_1^- \!-\! u_2^-$ \cite{MunganMert,Terzi}. For antiferromagnetic interactions, where
$c_{ij}\le 0$ and the flipping of hysteron $j$ from $0\!\rightarrow\! 1$ suppresses the flipping of hysteron $i$ from $0\!\rightarrow\! 1$, we obtain
three additional t-graphs which feature avalanches ($\{01\}\!\rightarrow\!\{10\}$) of length two (Fig.~\ref{Fig:4}(c)-\ref{Fig:4}(e)), while for purely ferromagnetic interactions ($c_{ij}\ge 0$), there are four additional t-graphs featuring avalanches $\{00\}\!\rightarrow\!\{11\}$ (Fig.~\ref{Fig:4}(f)-\ref{Fig:4}(i)). For
interactions of mixed sign, two additional t-graphs that both feature pseudo-avalanches of length three occur (Fig.~\ref{Fig:4}(j)-\ref{Fig:4}(k))--- e.g., the transition $\{00\}\!\rightarrow\!\{10\}$
in t-graph (j) proceeds via intermediate states
$\{01\}$ and $\{11\}$.
We note that for strong coupling, approximately $6$\% of  parameters
yield ill-defined  t-graphs (see Section \ref{sec:model}).
We conclude that for two hysterons, interactions yield qualitatively new transitions ---avalanches and pseudo avalanches--- and signficantly increase the number and diversity of t-graphs.

\subsection{Design inequalities for $n=2$ interacting hysterons}\label{sec:IIdes}

\begin{figure}[!t]
\begin{center}
\includegraphics[width=1\linewidth,bb = 72 104 808 468,clip]{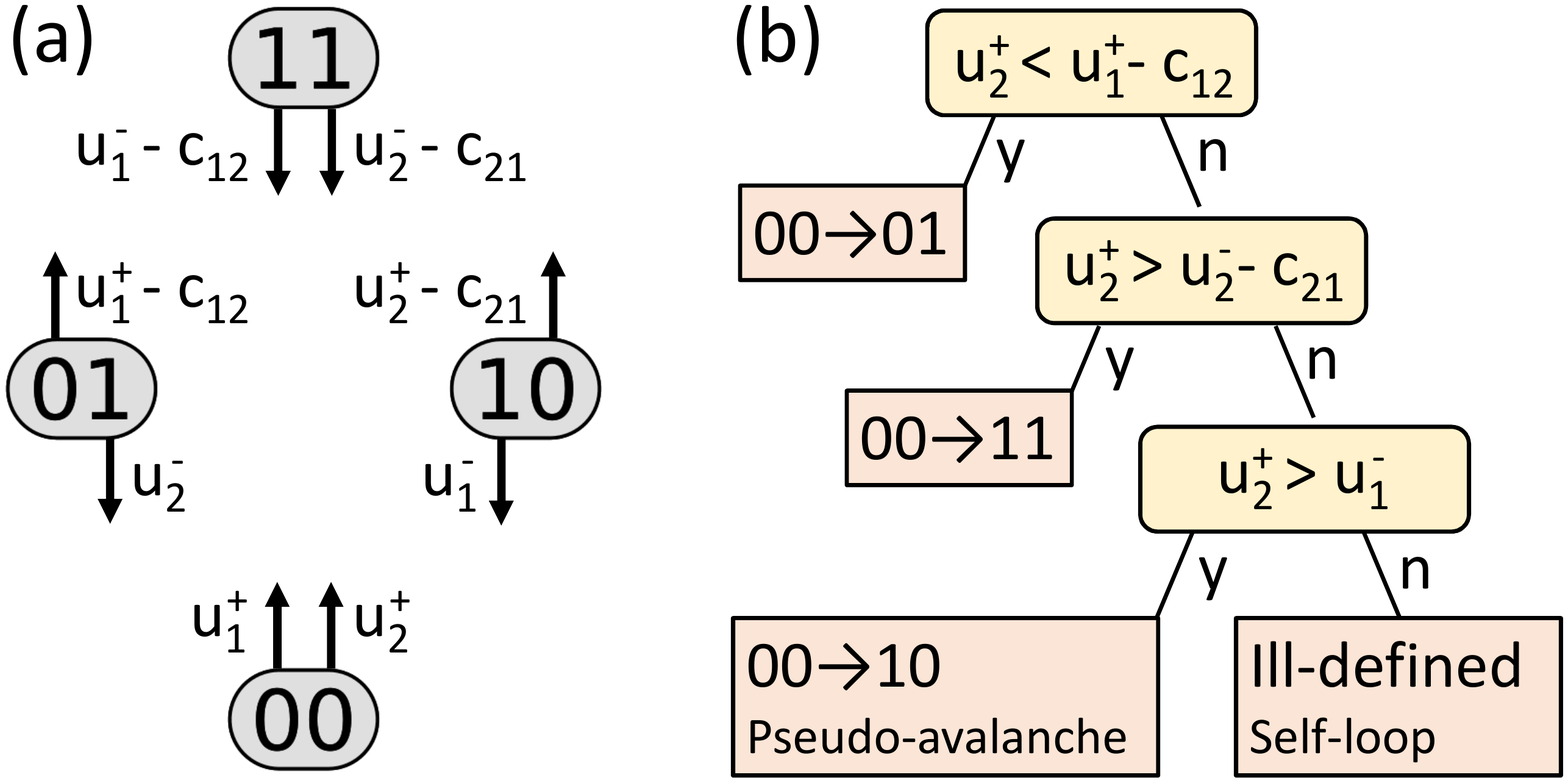}
\end{center}
\caption{(Color online) (a) Switching fields of each hysteron in each potential state of two interacting hysterons. Up and won arrows indicate up and down transitions; left and right positioning indicates switching fields for hysteron 1 and 2, respectively.
(b) Decision tree summarizing the conditions for transitions starting from state $\{00\}$. }
\label{Fig:5}
\end{figure}

One advantage of studying the case of only two interacting hysterons is that this allows to summarize the full set of conditions on the design parameters $u_i^\pm, c_{12}$ and $c_{21}$ for each of the 11 t-graphs of $n=2$ interacting hysterons.
These inequalities give insight into the nature of the design problem and the statistics as discussed in section \ref{sec:IIpwl}. Unfortunately, obtaining these inequalities is slightly tedious; after providing the general idea we focus
on a single subcase to guide the reader (Fig.~\ref{Fig:5}), before stating the final results in table \ref{table:overview}.

We recall the following design constraints on the bare switching fields:
\begin{eqnarray}
u_2^+ &<& u_1^+ ~,\label{d1}\\
u_1^- &<& u_1^+ ~,\label{d2}\\
u_2^- &<& u_2^+ ~.\label{d3}
\end{eqnarray}
To systematically determine the conditions for all potential $(2 \cdot 2^n-2)$ up and down transitions, we first determine the $n \cdot 2^n$ switching fields of each hysteron (Fig.~\ref{Fig:5}a). We then choose an initial state $S$ and  proceed via a three step process:
{\em (i)}: We determine the conditions for the
switching hysteron and the concommitant switching field $U^c(S)$.
(We note that while for $n=2$ only the down transition from state $\{11\}$
yields a non-trivial condition, for larger $n$ the situation is more elaborate.)
{\em (ii)}: We consider the stability conditions of the transition state $S'$ at $U^c$.
If $S'$ is stable, it is a landing state, and we have found an $S \rightarrow S'$ transition.
{\em (iii)}: When one stability condition for $S'$ is not met it is unstable, an avalanche occurs with a new switching hysteron and towards the next transition state $S'$ --- the critical U remains as determined in the first step. For this new state, we repeat step {\em (ii)-(iii)}. When more than one stability condition is not met, the avalanche is ill-defined. Similarly, when $S'=S$, the system is ill-defined as this causes a loop.

Starting out at state $S$, these three steps yield a set of inequalities for all transitions starting out at $S$ which span a decision tree.
We note that the number of inequalities quickly grows with the number of steps in an avalanche
(i.e., deeper into the three) and also with the number of hysterons, consistent with the explosive growth in the variety of t-graphs with $n$. Repeating this exercise for all states $S$, and collecting the inequalities in all corresponding threes,
while tedious, yields a complete set of necessary and sufficient conditions on the design parameters for each t-graph, as well as conditions for the t-graph to be ill-defined.

{\em Example.---}
To illustrate this approach, we explicitly consider the up transitions from state $\{00\}$ towards all other states following the three step process.
{\em(i)}: The design constraint Eq.~(\ref{d1}) determines that the 2nd hysteron will switch, so that $U^c=u_2^+$, and $S'=\{01\}$.
{\em (ii)} For  $S'$ to be stable at $U^c$, we require:
\begin{eqnarray}
U^c =u_2^+&<& u_1^+ -c_{12}~,\\
U^c =u_2^+ &>& u_2^-~,
\end{eqnarray}
where we note that the second condition is trivially satisfied due to the design constraint Eq.~(\ref{d3}).
Hence we can conclude the following necessary and sufficient condition:
\begin{equation}
\{00\}\rightarrow \{01\}: u_2^+<u_1^+ -c_{12}~. \label{00to01}\\
\end{equation}
We note that for non-interacting hysterons, where $c_{12}=0$, this condition is trivially satisfied due to design condition Eq.~(\ref{d1}). Moreover, this constraint can only be violated for positive $c_{12} > u_2^+ -u_1^+$. More generally, ''vertical'' and ''horizontal'' avalanches require positive and negative coupling coefficients.

We now iterate this process, by considering the case when the transition state $\{01\}$ is not stable, i.e. when Eq.~(\ref{00to01}) is not satisfied. We then obtain a new transition state $S'=\{11\}$, and we check the stability of
$S'$:
\begin{eqnarray}
u_2^+&>&u_1^- - c_{12}~,\label{00to11b}\\
u_2^+&>&u_2^- - c_{21}~.\label{00to11c}
\end{eqnarray}
We note that Eq.~(\ref{00to11b}) is automatically satisfied because of the design constraint Eq.~(\ref{d2}) and the fact that Eq.~(\ref{00to01}) is not satisfied. Inequality (\ref{00to11b}) is thus redundant and can be removed from our considerations - such dependencies between inequalities on the design parameters frequently occur. Hence, the necessary conditions for an avalanche from $\{00\}$ to $\{11\}$ are:
\begin{eqnarray}
\{00\}&\rightarrow& \{11\}: \nonumber\\
u_2^+&>&u_1^+ -c_{12}~, \label{00to11a} \\
u_2^+&>&u_2^- - c_{21}~.\label{00to11c}
\end{eqnarray}
We now consider
whether these equations are sufficient. For an avalanche, we do not keep track of the intermediate transition states and only monitor the initial and final state. In principle, the $\{00\}\rightarrow \{11\}$ avalanche might also have proceeded differently, e.g., via the $\{10\}$ state; in this specific case, the design constraint Eq.~\ref{d1} blocks this possibility. For larger systems, finding sufficient and necessary conditions for avalanches becomes much more involved. However, in this specific case,
Eqs.~(\ref{00to11a}) and (\ref{00to11c}) are both
sufficient and necessary conditions for the $\{00\}\rightarrow \{11\}$ avalanche.

We repeat this procedure again in case that
state $\{11\}$ is unstable at $U=u_2^+$, which happens when Eq.~(\ref{00to11a}) is satisfied, and Eq.~(\ref{00to11c}) is violated, and
which leads to a new transition state $\{10\}$.
Checking the stability of this state yields:
\begin{eqnarray}
u_2^+ &<& u_2^+ -c_{21}~, \label{10_stable_a}\\
u_2^+ &>& u_1^-~. \label{10_stable_b}\\
\end{eqnarray}
We note that $c_{21}$ has to be negative to violate
Eq.~(\ref{00to11c}), which implies that
Eq.~(\ref{10_stable_a}) is satisfied and can be removed from our consideration.
Hence, the conditions for the pseudo avalanche of length three $\{00\}\rightarrow \{10\}$ of the t-graph shown in Fig.~\ref{Fig:4}(j)  are:
\begin{eqnarray}
\{00\}&\rightarrow& \{10\}: \nonumber\\
u_2^+&>&u_1^+ -c_{12}~,  \\
u_2^+&<&u_2^- - c_{21}~,\\
u_2^+& >& u_1^-~ \label{ill}.
\end{eqnarray}

Finally, we note that violating the last equality Eq.~(\ref{ill}) yields a cycle $\{00\}\rightarrow\{00\}$, which yields the t-graph to be ill-defined; this indeed can happen (a concrete realization would be
$ u_1^+=0.5, u_1^-=0.8, u_2^+=1,u_2^-=0.7,c_{12}=0.1, c_{21} = -0.4$).
We summarize these findings in a decision three (Fig.~\ref{Fig:5}b), where we stress that for other states and for more hysterons, the situation generally is much more complex, featuring multiple conditions and outcomes per branch point, and cases where different branches yield the same transition.

\begin{figure}[!t]
\begin{center}
\includegraphics[width= \columnwidth,bb = 0 00 312 161,clip]{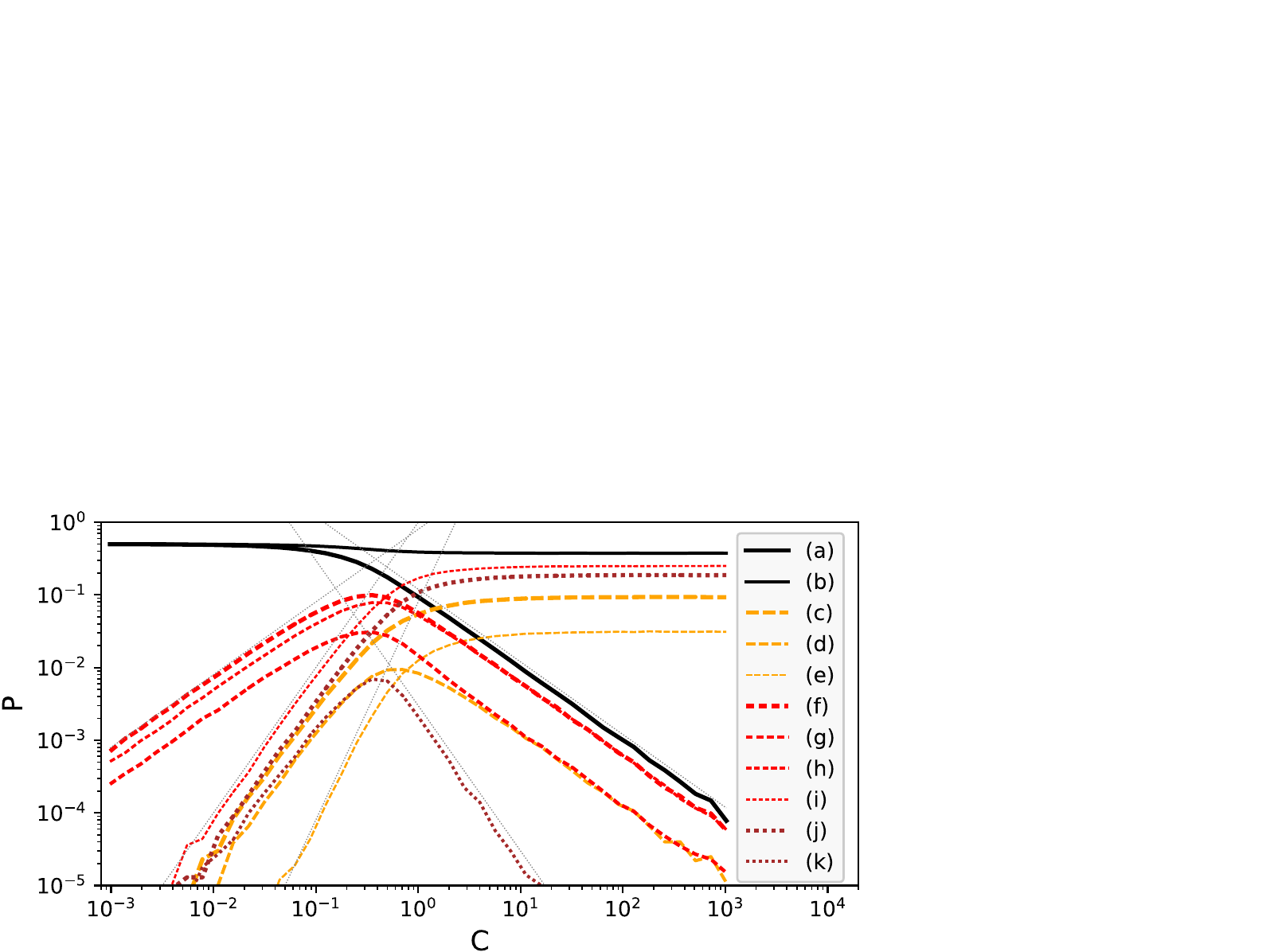}
\end{center}
\caption{(Color online) Probability of each t-graph shown in Fig.~\ref{Fig:4} as function of magnitude of the interactions $C$; fractions are for well-defined cases only (the fraction of ill-defined cases grows from 0\% at small $C$ to 6\% at large $C$). Thin grey
lines indicate integer powerlaw scalings.
}
\label{Fig:6}
\end{figure}

After collecting all inequalities for all possible $n=2$ transitions, and removing redundant inequalities, we obtain necessary and sufficient conditions for each of the 11 possible t-graphs,
as well as precise conditions for the occurrence of ill-defined graphs. We can express these by combining nine inequalities $x_i$, and we have checked these conditions numerically.
We note that these inequalities $x_i$ are not independent - for example, $x_1 \wedge x_8 \Rightarrow x_7$ - and we have checked that only 78 different combinations of $\{x_1,\dots,x_9\}$ arise.

The rather impenetrable nature of these sets of inequalities reflect the intrinsic complexity of the relation between design parameters and t-graphs. Nevertheless, we stress here that obtaining these inequalities is straightforward and could be automated. Solving these inequalities, i.e., finding design parameters for a given t-graph (potentially with additional constraints such as weak interactions, or with a number of interactions constants zero) can straightforwardly be implemented using linear programming. Finally, as we will see in the next section, these inequalities give crucial insight into the statistical properties of t-graphs.

\begin{table}[t]
\begin{tabular}{l|rrrrrrrr }
Condition~~ & \\

$x_1$& $-u_1^+$ & $      $ & $+u_2^+$ & $      $ & $+c_{12}$ & $       $ & $>$ & 0 \\
$x_2$& $      $ & $-u_1^-$ & $      $ & $+u_2^-$ & $+c_{12}$ & $       $ & $>$ & 0 \\
$x_3$& $      $ & $+u_1^-$ & $      $ & $-u_2^-$ & $-c_{12}$ & $+c_{21}$ & $>$ & 0 \\
$x_4$& $      $ & $+u_1^-$ & $      $ & $-u_2^-$ &           & $+c_{21}$ & $>$ & 0 \\
$x_5$& $      $ & $      $ & $-u_2^+$ & $+u_2^-$ & $       $ & $-c_{21}$ & $>$ & 0 \\
$x_6$& $+u_1^+$ & $      $ & $      $ & $-u_2^-$ & $-c_{12}$ & $+c_{21}$ & $>$ & 0 \\
$x_7$& $      $ & $-u_1^-$ & $+u_2^+$ &          &           &           & $>$ & 0 \\
$x_8$& $+u_1^+$ & $-u_1^-$ &          &          & $-c_{12}$ &           & $>$ & 0 \\
$x_9$& $      $ & $+u_1^-$ & $-u_2^+$ & $      $ & $-c_{12}$ &           & $>$ & 0 \\
\hline
\end{tabular}
\begin{tabular}{l|l}
t-graph~~ & ~~Condition\\
(a) &$~~ \neg {x_1} \wedge \neg x_3 \wedge \neg x_4 \wedge x_6 \wedge   x_7 $ \\
(b) &$~~  \neg x_1 \wedge  \neg x_2 \wedge  x_3 \wedge  x_6     $\\
(c) &$~~  \neg x_1  \wedge  \neg x_3 \wedge  \neg x_4 \wedge  \neg x_6 \wedge  x_7      \wedge  x_8 $\\
(d) &$~~   \neg x_1   \wedge  \neg x_3  \wedge  \neg x_4 \wedge  x_6  \wedge  \neg x_7     \wedge  x_8 $\\
(e) &$~~  \neg x_1  \wedge  \neg x_3    \wedge  \neg x_4  \wedge  \neg x_6 \wedge  \neg x_7  \wedge  x_8   $\\
(f) &$~~  \neg x_1  \wedge ((x_2  \wedge  x_3) \lor (\neg x_3  \wedge x_4   ))  \wedge   \neg x_5 $\\
(g) &$~~   x_1  \wedge  \neg x_2 \wedge  x_3  \wedge  \neg x_5  \wedge  x_6 \wedge  \neg x_9   $\\
(h) &$~~   x_1  \wedge  \neg x_3  \wedge  \neg x_4 \wedge  \neg x_5  \wedge  x_7 \wedge  \neg x_9   $ \\
(i) &$~~ x_1  \wedge  ((x_2  \wedge  x_3) \lor (\neg x_3  \wedge x_4 )) \wedge \neg x_5 $\\
(j) &$~~  x_1  \wedge \neg x_3  \wedge  \neg x_4  \wedge    x_5 \wedge  x_7   $\\
(k) &$~~ \neg x_1  \wedge  \neg x_3 \wedge  x_4  \wedge  x_5\wedge  x_6 $\\
ill & $ ~~ \neg x_3 \wedge x_5 \wedge \neg x_6 \wedge \neg x_7 \wedge \neg x_8$
\end{tabular}
\caption{Top: Nine conditions $x_i$ on the design parameters.
Bottom: Necessary and sufficient conditions for each $n=2$ t-graph (a)-(k) as well as for ill-defined t-graphs. }
\label{table:overview}
\end{table}

\subsection{Statistics}\label{sec:IIpwl}

We have sampled the probability for each t-graph to occur
as a function of $C$ (ensemble size $10^8$), for the 'mixed interaction' case where $|c_{ij}|\!\le\! C$ (Fig.~2(l)). While the interactions constants $c_{ij}$ are flatly sampled, only restricted by simple
independent constraints such as $|c_{ij}|<C$, the bare switching fields
have to satisfy two sets of constraints: First, we require
$u_i^- <u_i^+$ so that independent hysterons are well-defined, and second,
we require the ordering
of the upper switching fields ($u_1^+>u_2^+>\dots$) which limits the number of t-graphs by suppressing trivial permutations of the hysterons.
To numerically sample the switching fields that satisfy these constraints, we use an algorithm that guarantees that, for $c_{ij}\equiv 0$, all different orderings of the
lower switching fields, and thus all Preisach t-graphs, occur with equal probability.

As shown in Fig.~(\ref{Fig:6}), our data shows that all t-graphs can be realized for
arbitrary weak interactions. This can be understood from the invariance of the t-graph topology under shifts of the switching fields and multiplications of all parameters: The t-graph for
$\bar{c}_{ij}=\lambda c_{ij}$,
$\bar{u}_i^\pm= \lambda (u_i^\pm +1/\lambda) $
maintains its topology for arbitrarily small interaction constants ($\lambda\!\rightarrow \! 0$). Hence, for any set of parameters, we can find other parameters with arbitrarily small $c_{ij}$ such that the t-graph topology is maintained.
Crucially, this shows that weak interactions can break the Preisach phenomenology
when the switching fields are close to each other.

Moreover, we find that
these probabilities grow and decay as powerlaws $\sim \!C^n_i$ for small and large $C$, with integer exponents $n_i$. This is because some of the design inequalities are ''critical'' and require
the fine tuning of parameters when $c_{ij}$ is large or small.
As an example, consider condition $x_1:
u_2^+-u_1^+>-c_{12}$. As the design constraint Eq.~(\ref{d1}) stipulates that $u_2^+ < u_1^+$, $x_1$ can only be satisfied when $c_{12}$ is positive, and when $c_{12}\downarrow 0$ requires the difference between $u_1^+$ and $u_2^+$ to become vanishingly small, which statistically happens with probability ${\cal O} (|c|)$ - hence, $x_1$ is a critical condition for small $|c|$.
Similarly, conditions $x_5$, $\neg x_6$ and
$\neg x_8$ also are satisfied with probability ${\cal O} (|c|)$. Moreover, some combinations of condition may only occur with probability ${\cal O} (|c|)$. When $m$ independent critical conditions occur, t-graphs can only arise with probability ${\cal O} (|c|^m)$. Similarly, when $|c|$ becomes very large, some (combinations) of
the inequalities $x_i$ can only be satisfied when the coupling constants are of order one, which happens with probability ${\cal O} (|c|^{-1})$.
Together, these considerations explain the power law behavior seen in Fig.~\ref{Fig:6}, with
the number of independent critical conditions controlling $n_i$.

We finally note that while the details of the scaling of each t-graph may be intricate, the data in Fig.~\ref{Fig:6} suggests that understanding both the small
$C$ Preisach limit, as well as the large $C$ limit is sufficient to capture more of the trends. For large $C$, the interaction coefficients, $c_{12}$ and $c_{21}$ completely dominate all state switching fields $U_i^\pm$, with the bare switching fields acting as perturbations; while it's physical interpretation is not immediately clear, studying this limit in tandem with the small $C$ limit may provide insight into the statistical properties of (groups of) t-graphs.

\begin{figure}[!t]
\begin{center}
\includegraphics[width= 1\columnwidth,bb = 81 609 361 820,clip]{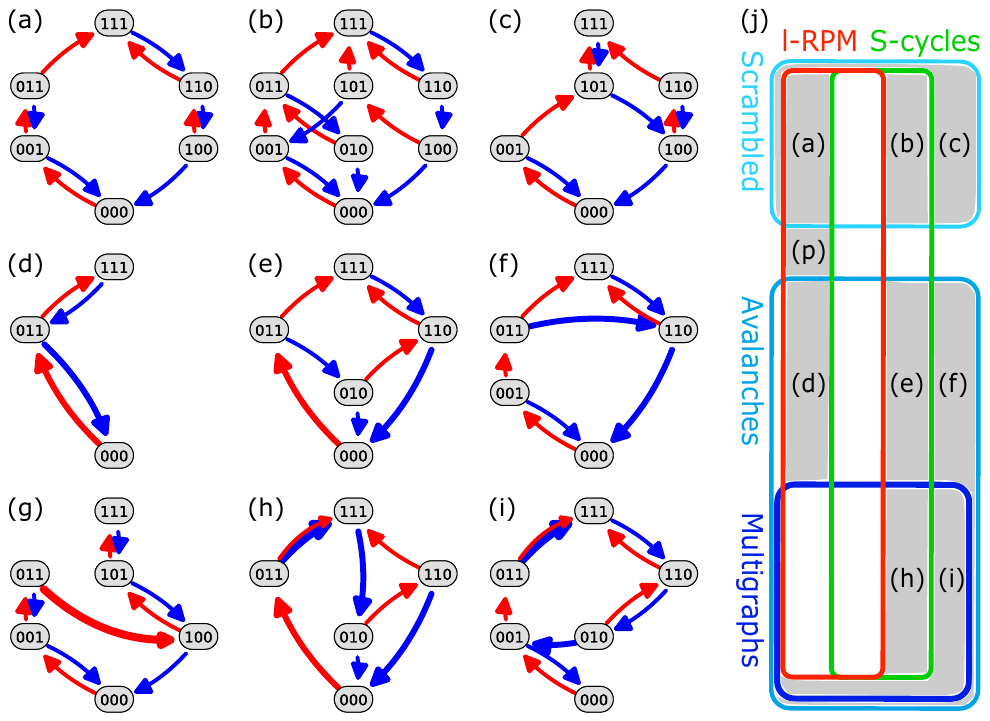}
\end{center}
\caption{(Color online) (a-c) Examples of scrambled t-graphs, featuring either l-RPM (a), a subharmonic-cycle (b), or neither (c).
(d-f) Examples of t-graphs with avalanches (thick arrows), featuring either l-RPM (d), a subharmonic-cycle (e), or neither (f). (g) t-graph featuring a dissonant avalanche. (h-i) Multi graphs with (h) and without (i) a subharmonic cycle.
(j) Venn-diagram for properties of $n\!=\!3$ t-graphs; letters refer to the examples in earlier panels, while Preisach t-graphs are indicated by '(p)'.
}
\label{fig:3}
\end{figure}

\begin{table*}[t]
\begin{tabular}{l|ccc|ccc|cccccc}
Panel & $u_1^+$ & $u_2^+$ & $u_3^+$ & $u_1^-$ & $u_1^-$ & $u_1^-$ & $c_{12}$ &
$c_{13}$ &$c_{21}$ & $c_{23}$ &
$c_{31}$ &
$c_{32}$ \\
\hline
(a) & 0.9 & 0.6 &0.5 &0.2 &0.4 &0.1 &0 &0 &0 &0 &-0.2 &-0.2 \\
(b) & 1 & 0.8 &0.6 &0.55 &0 &0.4 &-0.1 &-0.4 &-0.75 &-0.45 &-0.4 &-0.45 \\
(c) & 1 & 0.9 &0.8 &0.2 &0.8 &0 &0 &0 & 0 &-0.15 &-0.3 &0 \\
(d) & 1 & 0.7 &0.5 &0.3 &0.6 &0.4 &0.0 &-0.15 &0 &0.25 &0 &0 \\
(e) & 1 & 0.9 &0.8 &0.3 &0.2 &0.5 &0.25 &-0.2 &0 &0.2 &0 &0 \\
(f) & 0.9 & 0.7 &0.6 &0 &0.1 &0.3 &0.1 &-0.3 &0.3 &-0.25 &0 &-0.6 \\
(g) & 0.9 & 0.7 &0.5 &0.1 &0.4 &0.35 &0.3 &-0.4 &-0.7 &0 &-0.7 &0.1 \\
(h) & 1 & 0.95 & 0.9 &0.55 &0.6 &0.7 &0.23 &-0.3 &0.1  &0.4 & 0.3 & -0.18 \\
(i) & 0.75 & 0.6 &0.3 &0.35 &0.39 &0.28 &-0.05 &-0.30 &0 &-0.4 &0.05 &-0.6\\
\end{tabular}
\caption{
Examples of switching fields and coupling coefficients that produce the t-graphs shown in Fig.~\ref{fig:3}.
}
\label{table_fig3}
\end{table*}

\section{Three interacting hysterons}\label{sec:III}
The number and qualitative diversity of t-graphs mushrooms with $n$. We determined t-graphs for $10^8$ systems of
$n\!=\!3$ hysterons for $C\!=\!1$. We stress that in absence of interactions, there are only $n!=6$ distinct t-graphs \cite{MunganMert,Terzi}.
Strikingly, in the presence of interactions we obtain more than $15,000$ distinct
t-graphs.
To categorize the topology of these t-graphs, we focus on three characterizations: the nature of the transitions, the nature of pairs of transitions, and global topological measures of the t-graph. In particular,
individual transitions can feature avalanches, where more than one hysteron changes state simultaneously, and pairs of transitions can be scrambled, meaning that the switching order of hysterons becomes state dependent. These `local' features can lead to global t-graph topologies absent in the Preisach model, including
the break-down of loop-Return Point Memory, subharmonic cycles, and t-graphs that are multi-graphs. Together, these new features open up a large space of essentially unexplored behavior.

\subsection{Main features of t-graphs of coupled hysterons.}
To illustrate the main new features of t-graphs due to interactions,
we present nine examples of t-graphs that together illustrate the essential features (Fig.~\ref{fig:3}(a)-\ref{fig:3}(i)) (see Supplemental Information).

{\em (i) Scrambling.---}
In the Preisach model, the switching order is independent of state. We
define two non-avalanche transitions as scrambled when they are inconsistent with
such a state-independent ordering.
For example, consider the pair of
transitions $\{011\}\!\rightarrow\{001\}$ and
$\{111\}\!\rightarrow\{110\}$ in
the t-graph in Fig.~\ref{fig:3}(a).
The presence of the first transition implies that
$U_2^-(011) >U_3^-(011)$, and the second implies that
$U_2^-(111) <U_3^-(111)$. Such a pair of inequalities of the form
$U_i^\pm(S_1)>U_j^\pm(S_1)$ and
$U_i^\pm(S_2)<U_j^\pm(S_2)$
can only occur due to
hysteron interactions, and we label the pair of corresponding transitions as scrambled.
We note that when avalanches are present, their intermediate steps are ''hidden'', hindering to establish whether such a transitions are part of a scrambled pair of transitions---for the notion of scrambling, we therefore focus on pairs of direct (i.e., non-avalanche) transitions. As we will discuss in detail below, scrambling is a necessary, although not sufficient property for obtaining truly new behavior, and as such can be seen as the first step in a hierarchy of increasingly complex t-graphs that emerge due to interactions.

{\em (ii) Avalanches.---}
Without interactions, each transition corresponds to a single hysteron switching its phase, but in the presence of interactions many t-graphs
feature avalanches
where more than one hysteron changes phase simultaneously
(Fig.~\ref{fig:3}(d)-\ref{fig:3}(f)).
We note that ferromagnetic interactions promote ''vertical'' avalanches, where multiple hysterons collectively switch up or down. The magnetization $m({\bf s}):=\Sigma_i s_i$ then increases or decreases by more than one. In contrast, antiferromagnetic interactions promote ''horizontal'' avalanches, where the magnetization remains constant or changes at most by one. Mixed interactions
in addition can lead to more complex avalanches, such as the 'pseudo' avalanches shown in Fig.~(\ref{Fig:4}(j)-\ref{Fig:4}(k)).

{\em (iii) Dissonance.---}
Sofar, up and down transitions, initiated by an increase or decrease of the global driving $U$, lead to the
increase, respectively decrease of the magnetization $m$.
Remarkably, mixed ferro/antiferromagnetic interactions allow for {\em dissonant} avalanches, where an
up (down) avalanche leads to a decrease (increase) of the magnetization (Fig.~\ref{fig:3}(g)).

These three features significantly extend the space of possible t-graphs in comparison to those found in the Preisach model. Scrambling breaks the notion of a unique switching ordering, avalanches break the notion of nearby states, and dissonance blurs the connection between (in)decrease of the driving field, and
in(decrease) of the number of hysterons in state '1'. Collectively, these features lead to a range of new global behaviors of the the t-graphs:

{\em (iv) Multi-graphs.---} The presence of avalanches and dissonant transitions leads to cases where two states are connected both by an up and down transition---e.g., a pair of 'horizontal' avalanches that both preserve the magnetization, or an ordinary transition paired with a dissonant avalanche. The t-graphs then become directed {\em multigraphs}  (Fig.~\ref{fig:3}(h)-\ref{fig:3}(i)).
We stress here that it is essential for the algorithm that constructs t-graphs to allow for such multigraphs, which are surprisingly common for intermediate coupling coefficients.

{\em (v) Breakdown of loop-Return Point Memory.---}
Return Point Memory (RPM) occurs in a range of physical systems and in particular has been widely studied for the Preisach model \cite{Middleton,MunganMert,Terzi}. Loosely speaking, a system exhibits RPM when it revisits a previous state when the driving revisits a previous minimum or maximum of the driving.
A t-graph satisfies RPM when
one cannot escape a subloop without the driving passing through some previously established extremal values \cite{Terzi,MunganMert}.
While the presence of strict RPM may depend on the precise values of the switching fields \cite{Bense}, a recent definition of so-called loop-RPM (l-RPM)
focusses on the topology of the t-graph \cite{Terzi,MunganMert}. Essentially, l-RPM requires that each loop, given by a pair of 'top' and 'bottom' states connected by two sequences of purely up and a down transitions, is 'absorbing': this requires that any orbit starting from a state in this loop escapes the loop by going to either the top or bottom state, and not differently (see Supplemental Information for the precise definition). This definition is clearly analogous of that of RPM, and we note here that although RPM implies l-RPM, the converse is not necessarily true \cite{Bense}. The  t-graphs of the Preisach model all satisfy both properties, and
ferromagnetic  interactions have long been known to preserve RPM due to the so-called no-passing property \cite{Middleton,MunganMert,Terzi}.

In Fig.~4, only panels (a), (d) and (g) satisfy l-RPM:
(antiferromagnetic) interactions frequently break l-RPM. for example, in Fig.~\ref{fig:3}c the transitions  $\{100\}\!\rightarrow\!\{110\}\!\rightarrow\!\{111\}$ escape the
subloop between nodes $\{000\}$ and $\{101\}$. (in the SI we describe for all other t-graphs the precise transitions that break l-RPM). We finally note that
scrambling is a necessary, but not sufficient condition to break l-RPM.

{\em (vi) Subharmonic cycles.---}
Scrambling may also lead to subharmonic cycles (S-cycles),
where under cyclic driving the system
revisits earlier states only after more than one driving cycle. Similar to the discussion on l-RPM, we require here a definition in terms of the t-graphs topology, without regard to the precise switching values. Hence, we say the graph has an S-cycle if there are sequences of up and down transitions where one returns to the beginning state under more than one up/down subsequence.
In Fig.~\ref{fig:3}, panel (b), (e) and (h) show t-graphs with such S-cycles;
for example, the t-graph of Fig.~\ref{fig:3}(b) contains an S-cycle of period two:
$\{001\}\!\uparrow\!\{011\}\!\uparrow\!\{111\}\!\downarrow\!\{110\}\!\downarrow\!\{100\}\!\uparrow\!\{101\}\!\downarrow\!\{001\}\dots$
where $\uparrow$ and $\downarrow$ denote up and down transitions.

{\em (vii) Absence of l-RPM and S-cycles}
While we observe that l-RPM and the presence of S-cycles are mutually exclusive for $n=3$, we stress here that it is also possible to break l-RPM without having a S-cycle, as shown in Fig.~\ref{fig:3}(c), (f) and (i).

Scrambling is a necessary, but clearly not sufficient condition to break l-RPM or obtain S-cycles. Beyond that, we find that, at least for $n=3$, the ``local'' measures---scrambling, avalanches, multigraphs---can occur concurrently with the ``global'' measures, l-RPM and the presence of S-cycles, except that
multi-graphs can never satisfy l-RPM (Fig.~\ref{fig:3}(j))
This can easily be understood by noting that a multi-edge in a given loop implies that an up-boundary contains a down transition, or a down-boundary contains an up-transition, which allows to establish an orbit that violates l-RPM.

In table \ref{table_fig3} we present examples of switching fields and coupling coefficients that produce t-graphs with the same topologies as shown in Fig.~\ref{fig:3}. These parameter values have been selected after some manual optimization steps, setting some small interactions to zero and rounding of all values to at most two significant digits. While these parameters are not optimal in any well-defined sense, they may serve as specific starting points for further studies, as well as to guide the reader in the construction of t-graphs by providing specific examples. Moreover, we have numerically checked that for these parameters, small changes of ${\cal O}(10^{-3}$) do not change the topology of the t-graph, thus demonstrating that even rare graphs are robust.

\begin{figure}[!t]
\begin{center}
\includegraphics[width= 1\columnwidth,bb = 81 450 361 610,clip]{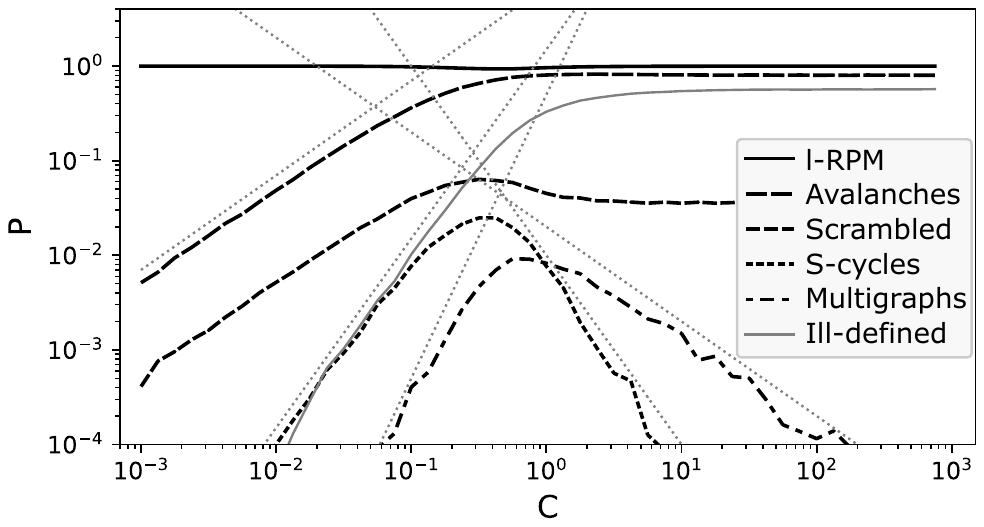}
\end{center}
\caption{Probabilities of $n\!=\!3$ t-graph types, where $c_{ij} \in [-C,C]$.
These probabilities grow and decay as integer power-laws (dashed lines).
While scrambling, S-cycles, and multigraphs most likely arise for intermediate interactions, the fraction of
t-graphs featuring avalanches increases with $C$ and plateaus at 80 \%.
The fraction of parameters that yield
ill-defined t-graphs also increases with $C$ and plateaus at 57 \%;
fractions are for well-defined cases only.
}
\label{fig:3b}
\end{figure}

In summary, these examples illustrate how hysteron interactions generate a host of new features of the t-graphs. In particular,
scrambling breaks the state independent ordering of transitions seen in Preisach t-graphs, avalanches and dissonance
enlarge the types of transitions between states, and together these can yield multi-graphs, breakdown of l-RPM and subharmonic cycles.


\subsection{Statistics}

For $C=1$, more that 62\% of distinct t-graphs break l-RPM. However, not all of these t-graphs are statistically equally likely. To probe the statistical properties,
we have sampled the probability of l-RPM, avalanches, scrambling, S-cycles, multi-graphs and ill-defined cases as function of $C$ for an ensemble size $10^5$ (Fig.~\ref{fig:3b}).
Strikingly, the majority of random parameters yield t-graphs that satisfy l-RPM
(minimum fraction $\sim 0.93$ for $C\approx 0.4$).  Hence, while a
fraction of all t-graphs dominates the statistics, interactions produce
a wide variety of t-graphs.

The probabilities of each class of t-graphs vary similarly to the probability of individual t-graphs, with integer powerlaws $\sim C^n$, and also point to well defined behavior in the $C\rightarrow \infty$ limit. Most interesting behavior occurs for $C$ between 0.1 and 1, where the probability of scrambled transitions, S-cycles and multigraphs peaks. For large $C$ the number of ill-defined t-graphs plateaus at 57\%, and explorations for larger $n$ indicate even larger percentages. This suggests that additional rules that avoid ill-defined transitions and/or loops are necessary to study the behavior of larger systems with strong interactions. Finally, for the remaining 43\% of parameters that yield well-defined t-graphs, most lead to t-graphs with avalanches and which satisfy l-RPM.

\begin{figure}[!t]
\includegraphics[width=\linewidth,bb = 000 00 390 170]{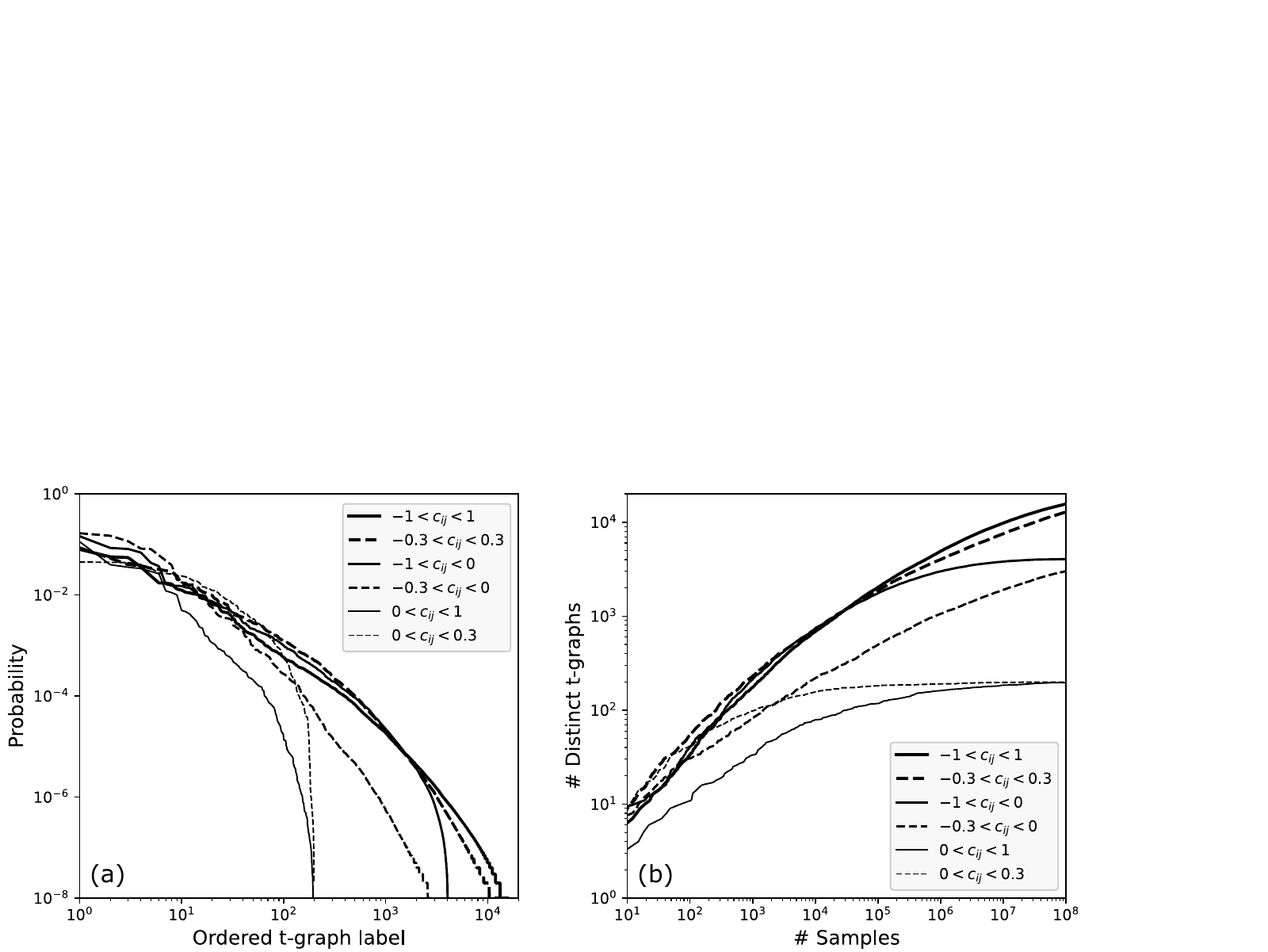}
\caption{Statistical measures of $n=3$ t-graphs, for two different interactions strengths
and for mixed, ferromagnetic, and purely anti-ferromagnetic interactions. (a) Individual t-graphs occur with probabilities that can span at least four orders of magnitude. (b) As a function of the number of realizations, the number of t-graphs grows slowly, and for mixed interactions has not saturated for $10^8$ samples.	 }\label{fig_n3_stat}
\end{figure}

We have further explored differences between purely ferromagnetic, purely antiferromagnetic, and mixed interactions. First, we demonstrate that the statistical weight of individual t-graphs is broadly distributed, by studying $10^8$ t-graphs realized for interactions strengths $C=0.3$ and $C=1$
and for mixed ($-C\!<\!c_{ij}\!<\!C$), ferromagnetic ($0\!<\!c_{ij}\!<\!C$), and purely anti-ferromagnetic ($-C\!<\!c_{ij}\!<\!0$) interactions.
By ordering each t-graph by its probability (from high to low),
we observe that the
probabilities for a given t-graph span many decades, with the majority of t-graphs
spanning a small fraction of parameter space
(Fig.~\ref{fig_n3_stat}(a)).

Second, and consistent with the profusion of 'rare t-graphs, we find that the number of distinct t-graphs as function of the number of samples grows slowly (Fig.~\ref{fig_n3_stat}(b)).
In particular, while for $10^8$ realizations the number of t-graphs for purely ferromagnetic and antiferromagnetic appears has (nearly) saturated around 198 and ${\cal O}(4050)$, the number of t-graphs for mixed interactions is still growing (Fig.~\ref{fig_n3_stat}(b)). Hence, exhaustive sampling, or smarter techniques to map out the space of t-graphs can be expected to yield even more, rare t-graphs.

Third, we observe purely ferromagnetic interactions  do not generate S-cycles, multigraphs or dissonant avalanches and only produce t-graphs that satisfy l-RPM (possibly with ''vertical'' avalanches). This is completely consistent with earlier observations that ferromagnetic interactions preserve the no-passing property, severely restricting the t-graphs and pathways  \cite{RMP,Barker,Middleton,Sethna,Terzi,MunganMert}.
In contrast, purely antiferromagnetic can break l-RPM, generate S-cycles, and yield multi-graphs, but cannot create dissonant avalanches.
Together, this shows that while antiferromagnetic interactions are essential to obtain exotic behavior, mixed interactions produce the largest variety of t-graphs.

Together, interacting hysterons lead to a large space of essentially unexplored t-graphs. Three features stand out.
First, while statistically, l-RPM is the most likely global behavior, even for strong coupling, a large number of t-graphs with qualitatively different features can be found. Second, scrambling breaks the unique ordering in the switching sequences, and
is a necessary, although not sufficient property to break l-RPM, and as such can be seen as the first step in a hierarchy of increasing complexity---in contrast, ordinary avalanches seem to have less of an impact on the global features of t-graphs. Third, dissonant avalanches break the link between up/down transitions in in/decrease of the magnetization, and open up the possibility of multi-graphs, which never satisfy l-RPM and lead to even more strongly non-classical behavior.

\section{Designer Pathways}
We suggest that the complex pathways of interacting hysterons naturally can be described in the language of computing. In particular, the directed graphs that encode sequential
computations in finite state machines \cite{FSM} are strongly reminiscent of
t-graphs, where the labels of each edge (''up transition at $U=0.5$'') play the role of the input to the ``hysteron machine''.

\begin{figure}[!t]
\begin{center}
\includegraphics[width= \columnwidth,bb = 80 527 236 663,clip]{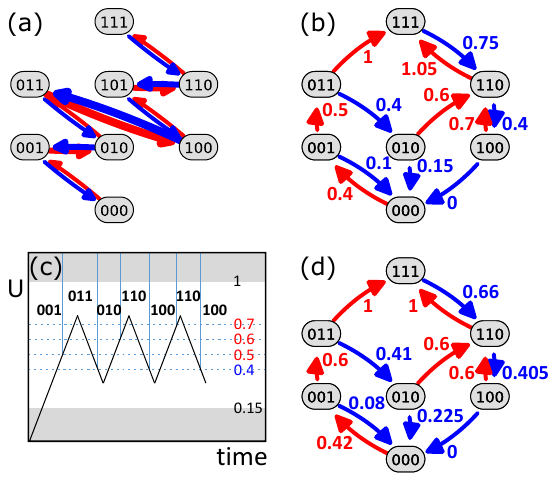}
\end{center}
\caption{(Color online) (a) T-graph that sequentially accesses  states by sweeping $U$ (we have changed the curvature of the arrows for visibility).
(b) Accumulator t-graph with switching fields as indicated (Table III(b)). (c) Response for the t-graphs shown in panel (b), demonstrating that
for $0.7\!<\!U_M\!<\!1$, states $\{110\}$ and $\{100\}$ are only reached after two driving cycles. Dashed lines indicate critical switching fields, and grey regions indicate potential transitions to the extremal states. (d) Switching fields corresponding to an alternative design for the accumulator
(Table III(d)).
}
\label{fig:4}
\end{figure}

As a first example of such a hysteron machine we exploit dissonant avalanches to realize t-graphs that contain all eight states in a single pathway of up (or down) transitions. In our dataset, 740 realizations representing 51 distinct t-graphs contain such pathways. We select an example where both the up and down pathways between ${\bf 0}$ and ${\bf 1}$ follow the ordered binary numbers 000-111, and which acts as an analog-digital converter (ADC; Fig.~\ref{fig:4}(a)).
The design inequalities specify a linear programming problem \cite{PreprintKeim}, and a judicious choice of parameters (see table \ref{table_fig4})
allows to
tune the critical switching fields of the seven up and seven down transitions exactly to values $0.1,0.2,\dots,0.7$, respectively $0.65, 0.55, \dots, 0.05$, making all states easily addressable and all transitions between states hysteretic (as required for ADCs). We have in addition verified that for the this design the t-graph's topology is stable to random perturbations of the design parameters of at least magnitude $10^{-3}$.

\begin{table*}[t]
\begin{tabular}{l|ccc|ccc|cccccc}
Panel & $u_1^+$ & $u_2^+$ & $u_3^+$ & $u_1^-$ & $u_1^-$ & $u_1^-$ & $c_{12}$ &
$c_{13}$ &$c_{21}$ & $c_{23}$ &
$c_{31}$ &
$c_{32}$ \\
\hline
(a) & 0.8 & 0.4 &0.1 &0.35 &0.15 &0.05 &0 &0 &0 &0 &-0.2 &-0.2 \\
(b) & 0.7 & 0.45 &0.4 &0 &0.15 &0.1 &0.1 &-0.4 &-0.25 &-0.05 &-0.35 &-0.3 \\
(d) & 0.7 & 0.45 &0.42 &0 &0.255 &0.08 &0.1 &-0.4 &-0.15 0 &-0.15 &-0.25 &-0.33 \\
\end{tabular}
\caption{
Examples of switching fields and coupling coefficients that produce the t-graphs shown in Fig.~\ref{fig:4}.
}
\label{table_fig4}
\end{table*}

As a second example, we explore the breakdown of l-RPM and select an 'accumulator' t-graph that contains the pathway $\{001\}\!\uparrow\!\{011\}\!\downarrow\!\{010\}\!\uparrow\!\{110\}\!\updownarrow\!\{100\}$ (Fig.~4(b)-(c)). Hence, under cyclic driving, a system described by such a t-graphs 'counts to two'. This behavior has recently been observed by us in experiments, and may be related to transient memory \cite{bense,PreprintLindeman}.

This example is particularly rich, as even for a given topology, different response can be encoded depending on the precise values and orderings of the switching field.
To demonstrate this, we first
choose parameters such that the switching fields of the two down transitions
$\{011\}\!\downarrow\!\{010\}$ and
 $\{110\}\!\downarrow\!\{100\}$ are equal to 0.4, while the switching fields for
 $\{001\}\!\uparrow\!\{011\}$,
$\{010\}\!\uparrow\!\{110\}$, and
$\{100\}\!\uparrow\!\{110\}$ are equal to 0.5, 0.6 and 0.7, respectively (Table.~\ref{table_fig4}(b)).
We have verified that the t-graph's topology is stable to random perturbations of the design parameters of magnitude $10^{-3}$.

The response of this system when $U$ is  cycled between $u_m\!>\!0.15$
and $u_M<1$ evidences different accumulator/counting behavior. For $0.7\!<\!U_M\!<\!1$, the system reaches state
$\{011\}$ at the first peak, and state $\{110\}$ at subsequent peaks:  this pathway distinguishes between one or more cyclical drivings (Fig.~\ref{fig:4}c).
Moreover, for $0.6\!<\!U_M\!<\!0.7$, the first cycle reaches $\{011\}$, the second $\{110\}$ and subsequent cycles remain stuck at $\{100\}$ (''counting to three''); for $0.5\!<\!U_M\!<\!0.6$, the first cycle reaches  $\{011\}$ and subsequent cycles remain stuck at $\{010\}$. Hence, a collection of three hysterons with appropriate interactions and switching fields can accumulate/count to two or three, depending on driving amplitude.

The parameters in our model offer freedom in the choice of the critical switching fields, although there are some constraints. For example, requiring that all three relevant up transitions in the accumulator t-graph ( $\{001\}\!\uparrow\!\{011\}$,
$\{010\}\!\uparrow\!\{110\}$,
$\{100\}\!\uparrow\!\{110\}$) are equal, necessitates the two down transitions
($\{011\}\!\downarrow\!\{010\}$,
$\{110\}\!\downarrow\!\{100\}$) to be unequal. To see this, we notice that in terms of the design parameters, the three up transitions are at $u_2^+-c_{23}, u_1^+-c_{12}$ and $u_2^+-c_{21}$ respectively, so that when all are equal, $c_{21}=c_{23}$. Similarly,
for the  down transitions to have equal switching fields, we require $u_3^- -c_{32}=u_2^- -c_{21}$. Finally, for the down transition from state $\{011\}$ land on state $\{010\}$ and not on $\{001\}$, we require the design inequality $u_3^- - c_{32}> u_2^- - c_{23}$.
When the pair of down transitions are equal,
this latter inequality can be rewritten as $u_2^- -c_{21} > u_2^- - c_{23} \Rightarrow c_{21} < c_{23}$, which is in disagreement with the requirement that all three up transitions are at the same switching field.

Nothwithstanding this constraint, it is easy to find design parameters so that the three up-transitions are equal (Table.~\ref{table_fig4}(d)), yielding the critical switching fields shown in Fig.~\ref{fig:4}(d). As all the relevant up switching fields are equal, the only counting behavior that is left is ''counting to two''. This example demonstrates that even for a given t-graph topology, qualitatively
distinct responses and finite state machines may be encoded.

We suggest that a wide variety of more complex functions may be achievable in t-graphs that encode different topologies, such as (longer) S-cycles, multiple S-cycles, etc. Paired with the design options that e.g., tune all relevant up transition fields to the same value or specifically ordered values, the design space for complex hysteron ''machines'' is very large.

\vspace{5mm}


\section{Discussion} This work highlights that even small collections of weakly interacting hysterons exhibit a staggering multitude and variety of pathways and t-graphs, and  suggests that
hysterons with appropriate thresholds and interactions can act as information processing devices.
We highlight a number of key questions.
First, the types of t-graphs and underlying computations that can be realized
by interacting hysterons is unknown, with interesting sub-questions arising for  interactions that are purely ferromagnetic, purely antiferromagnetic, reciprocal ($c_{ij}\!=\!c_{ji}$ \cite{PreprintKeim}), or sparse.
 Second,
in exploratory studies we have found that the fraction of random parameters that yield ill-defined t-graphs increases with $n$ and $C$ and asymptotes to one. This suggests that strongly coupled systems cannot trivially be described by hysterons, and more advanced models, that avoid ambiguities due to multiple unstable hysterons or self-loops, are called for.
Third, metamaterials might yield
physical realizations, with serially coupled mechanical hysterons naturally implementing  anti-ferromagnetic interactions \cite{PreprintZanaty,PreprintKaren,in_prep}.
Finally, viscoelastic effects could be leveraged to obtain rate-dependent pathways and t-graphs and self-learning systems \cite{rate_dep_meta1,
rate_dep_meta2,
rate_dep_meta3, learning_plasticity1,learning_plasticity2}.
Together, progress on these questions will realize  targeted
pathways and information processing in designer materials.


{\em Acknowledgements.---}
We acknowledge insightful discussions with H. Bense, K. Bertoldi, N. Keim, G. Korovin, Y. Lahini, C. Lindeman, M. Mungan, S. Nagel, J. Paulsen and M. Zanaty.

\section{SI}

\subsection{l-RPM}
To check whether a t-graph has the property of l-RPM, we first
define operators $U$ and $D$ so that $U(S)$ (and $D(S)$)
produces the state after an up (down) transition from state $S$, and
define the full up orbit $Up(S)$
as the sequence of states $U(s), U^2(s), \dots$, and similarly for the down orbit $Dw(S)$.
Consider a loop, defined by a pair of states $S_m$ and $S_M$, such that the system evolves from $S_m$ to $S_M$ (and vice-versa) by a series of exclusively up (down) transitions.
Define the up boundary of a loop as the sequence of states obtained by repeatedly applying $U$ on $S_m$ until $S_M$ is reached, and the down boundary as the sequence of states by applying
$D$ on $S_M$ until $S_m$ is reached.
A loop is said to be absorbing if the states $S_m$ ($S_M$) are part of the down (up) orbit from any state which is part of the up (down) boundary---this implies that the system can only escape a subloop through the extremal states.
A t-graph has the l-RPM property if and only if all loops are absorbing \cite{Terzi,MunganMert}.

\subsection{Properties and pathways t-graphs in Fig.~7.}
Here we summarize the specific properties and pathways (in particular, S-cycles and orbits that break l-RPM) of the t-graphs in Fig.~7.

Fig.~7(a): 
This t-graph has no cycles and satisfies l-RPM; it does contain scrambled transitions.

Fig.~7(b): 
This t-graph has a subharmonic cycle:
$\{001\}\!\uparrow\!\{011\}\!\uparrow\!\{111\}\!\downarrow\!\{110\}\!\downarrow\!\{100\}\!\uparrow\!\{101\}\!\downarrow\!\{001\}\dots$
where $\uparrow$ and $\downarrow$ denote up and down transitions.
Moreover, it breaks l-RPM. To see this,
consider the sub-loop between nodes
$S_m\!=\!\{100\}$ and $S_M\!=\!\{111\}$. Node $\{101\}$ is part of the up boundary. However, its down orbit, $\{101\}\!\rightarrow\!\{001\}\!\rightarrow\!\{000\}$ never reaches its $S_m$. Hence, this loop is not absorbing, and the t-graphs does not have the l-RPM property. In more compact notation:
$\left[S_m,S_M\right]\!=\!\left[ \{100\},\{111\}\right]$; $\{101\}\in$ up boundary; $S_m \notin Dw(\{101\})$.


Fig.~7(c): 
This t-graphs does not have an S-cycle, yet breaks l-RPM:
$\left[S_m,S_M\right]\!=\!\left[ \{000\},\{101\}\right]$; $\{100\}\in$ down boundary; $S_M \notin Up(\{100\})$.

Fig.~7(d): 
This t-graph contains two avalanches of length two, no S-cycles, and satisfies l-RPM.

Fig.~7(e): 
This t-graph has avalanches and an S-cycle:
$\{000\}\!\uparrow\!\{011\}\!\downarrow\!\{010\}\!\uparrow\!\{110\}\!\downarrow\!\{000\}
\!\dots$. It also breaks l-RPM:
$\left[S_m,S_M\right]\!=\!\left[ \{000\},\{011\}\right]$; $\{010\}\in$ down boundary; $S_M \notin Up(\{010\})$.

Fig.~7(f):
This t-graph contains avalanches, no S-cycles, yet breaks l-RPM:
$\left[S_m,S_M\right]\!=\!\left[ \{000\},\{011\}\right]$; $\{110\}\in$ down boundary; $S_M \notin Up(\{110\})$.


Fig.~7(g): 
This t-graph has no cycles and satisfies l-RPM. We note that the simultaneous presence of the transitions
$\{001\}\!\rightarrow\!\{011\}$ and $\{100\}\!\rightarrow\!\{101\}$ indicates scrambling.

Fig.~7(h): 
This t-graph is a multigraph and contains a S-cycle:
$\{000\}\!\uparrow\!\{011\}\!\uparrow\!\{111\}\!\downarrow\!\{010\}\!\uparrow\!\{110\}\!\downarrow\!\{000\}\!\dots$.
Moreover, this t-graph contains several orbits that violate l-RPM.
For sub-loop $\left[S_m,S_M\right]\!=\!\left[ \{000\},\{011\}\right]$,
states $\{111\}$ and $\{010\}$ are part of the down boundary, but neither of their up orbits reach $S_M$; moreover for sub-loop  $\left[S_m,S_M\right]\!=\!\left[ \{010\},\{111\}\right]$,
state $\{110\}\in$ up boundary, while
$S_m \notin Dw(\{110\})$.

Fig.~7(i): 
This t-graph is a multigraph, has no S-cycles yet many examples of orbits that violate l-RPM. For
the sub-loop
$\left[S_m,S_M\right]\!=\!\left[ \{000\},\{011\}\right]$,
states $\{111\}$, $\{110\}$ and $\{010\}$ are parts of the down boundary, yet their up orbits do not reach $S_M$. Moreover, also for sub-loop $\left[S_m,S_M\right]\!=\!\left[ \{001\},\{011\}\right]$, states $\{111\}$, $\{110\}$ and $\{010\}$ are parts of the down boundary, yet none of their up orbits reach $S_M$.

\end{document}